\documentclass[11pt,a4paper]{article}

\usepackage[utf8]{inputenc}
\usepackage[T1]{fontenc}
\usepackage[polish,english]{babel}
\usepackage{amsmath,amssymb,amsthm}
\usepackage{graphicx}
\usepackage{geometry}
\usepackage[backend=biber,style=authoryear,natbib=true]{biblatex}
\usepackage{hyperref}
\usepackage{orcidlink}

\newcommand{\dd}{\mathrm{d}}

\newcommand{\Var}{\operatorname{Var}}
\newcommand{\e}{\mathrm{e}}
\newcommand{\ii}{\mathrm{i}}
\newcommand{\sx}{\hat{\sigma}_x}

\newcommand{\sz}{\hat{\sigma}_z}
\newcommand{\Ham}{\hat H}
\newcommand{\ddt}{\frac{\dd}{\dd t}}
\newcommand{\ssx}{\langle\sx\rangle}
\newcommand{\keyword}[1]{\par\medskip\noindent\textbf{Keywords:} #1}
\newcommand{\funding}[1]{\par\medskip\noindent\textbf{Funding:} #1}
\newcommand{\dataavailability}[1]{\par\medskip\noindent\textbf{Data Availability Statement:} #1}

\title{Operational Meaning of the Mandelstam--Tamm Time Scale in a Spin-$1/2$ Clock\\
	\small A Comment on Georgiev, \emph{Symmetry} \textbf{16}, 100 (2024)}
\author{
	Jakub (James) Rog\,\\[4pt] 
	{\small Faculty of Philosophy, University of Warsaw; Krakowskie Przedmieście 3, Warsaw, Poland}\\
	{\small Correspondence: \href{mailto:j.rog2@uw.edu.pl}{j.rog2@uw.edu.pl}}
}
\date{\today}

\begin{document}
\maketitle

\begin{abstract}
	Georgiev's analysis of a freely precessing spin correctly emphasizes that the Mandelstam--Tamm quantity
	$\Delta T=\Delta\hat A/|\dd\langle\hat A\rangle/\dd t|$ is not the variance of a time operator. He concludes, however, that its divergence near the extrema of the clock signal is devoid of physical meaning and that $\Delta T$ does not disclose how well the system measures the passage of time. For the same projective $\hat\sigma_x$ readout, we show that $\Delta T$ equals the inverse square root of the classical Fisher information in one binary outcome at regular operating points. Its divergence at submaximal-coherence extrema therefore identifies a loss of local sensitivity in the selected readout, not an infinite variance of physical time; a complementary $\hat\sigma_y$ measurement remains informative. This interpretation also separates Georgiev's exactly known tick--tock period from the statistical precision with which a tick can be located. It preserves his warning against a time-operator interpretation while qualifying his stronger conclusion that the Mandelstam--Tamm scale lacks physical content.
\end{abstract}

\keyword{Mandelstam--Tamm relation; time estimation; Fisher information; spin clock; quantum measurement}

\section{Scope of the comment}

Georgiev usefully distinguishes the external time parameter from a time operator. He considers the standard Mandelstam--Tamm relation \cite{MandelstamTamm1945}, and we agree that this is not a Robertson relation between time and energy as canonical observables. Our target is Georgiev's narrower inference: he takes the time dependence of the Mandelstam--Tamm scale $\Delta T$ to disqualify it as a measure of clock quality, and he contrasts its divergence at the extrema of the clock signal with the exactly known interval between consecutive extrema to conclude that it is not a meaningful representation of time.

Georgiev derives the binary outcome probabilities of the spin readout, but does not treat them as a statistical model for estimating $t$ or consider their Fisher information. This omission matters because asking how well a clock measures time is operationally asking how well its outcomes distinguish nearby values of $t$; Fisher information is the standard local measure of that distinction and controls estimator precision through the Cram\'er--Rao bound \cite{Paris2009}. We therefore read Georgiev's probabilities as a time-parameterized model, identify $\Delta T$ with its inverse Fisher sensitivity, compare the selected readout with the quantum information fixed by $\Delta E$, and then revisit his divergence and tick--tock arguments. This route does not promote time to an observable; it supplies the missing readout-level meaning.

\section{The spin clock and the disputed inference}
\label{sec:model}

Georgiev considers the following Hamiltonian, initial state, normalization, and coherence parameter in his Eqs.~(20), (23), and (31):
\begin{equation*}
  \Ham=\frac{\hbar\omega}{2}\sz,
  \qquad
  |\Psi(0)\rangle=\alpha_1|{\uparrow_z}\rangle
  +\alpha_2|{\downarrow_z}\rangle,
  \qquad
  C=2|\alpha_1||\alpha_2|,
  \label{eq:model}
\end{equation*}
where $|\alpha_1|^2+|\alpha_2|^2=1$ and $C$ is the $\ell_1$ norm of energy coherence. For this pure qubit, $C$ is also the transverse Bloch-vector length. Georgiev's Eqs.~(41) and (29) give
\begin{equation*}
 \langle\sx\rangle
 =2\!\left[
 \operatorname{Re}(\alpha_1\alpha_2^*)\cos(\omega t)
 +\operatorname{Im}(\alpha_1\alpha_2^*)\sin(\omega t)
 \right],
 \label{eq:Georgiev-signal}
\end{equation*}
and
\begin{equation*}
  \Delta\Ham=\frac{1}{2}\hbar\omega C.
  \label{eq:deltaH}
\end{equation*}

Substitution of his Eqs.~(42) and (44) into the definition in Eq.~(45) yields Georgiev's Eq.~(46):
\begin{equation*}
 \Delta T
 =\frac{\Delta\sx}
 {\left|\frac{\dd}{\dd t}\langle\sx\rangle\right|}
 =\frac{
 \left|\alpha_1^2\e^{-\ii\omega t}
 -\alpha_2^2\e^{\ii\omega t}\right|}
 {2\omega\left|
 \operatorname{Re}(\alpha_1\alpha_2^*)\sin(\omega t)
 -\operatorname{Im}(\alpha_1\alpha_2^*)\cos(\omega t)
 \right|}.
 \label{eq:DT-Georgiev}
\end{equation*}

Here $\omega>0$ denotes the angular frequency. For convenience let
\begin{equation*}
 \alpha_1\alpha_2^*
 =|\alpha_1\alpha_2|\e^{\ii\phi},
 \qquad
 \vartheta=\omega t-\phi.
 \label{eq:phase-shorthand}
\end{equation*}

Equations~(42)--(44) of Ref.~\cite{Georgiev2024} then give
\begin{align*}
 \Delta\sx
 &=
 \left|\alpha_1^2\e^{-\ii\omega t}
 -\alpha_2^2\e^{\ii\omega t}\right|
 =\sqrt{1-\ssx^2}
 =\sqrt{1-C^2\cos^2\vartheta},
 \notag\\
 \langle\sx\rangle&=C\cos\vartheta,
 \\
 \left|\frac{\dd}{\dd t}\langle\sx\rangle\right|
 &=\omega C|\sin\vartheta|.
 \label{eq:DT-ingredients}
\end{align*}
Note that the time argument of $\ssx$ is suppressed in the
notation, although $\ssx$ remains time dependent.

Hence
\begin{equation}
 \Delta T
 =\frac{\sqrt{1-C^2\cos^2\vartheta}}
 {\omega C|\sin\vartheta|}.
 \label{eq:DT-compact}
\end{equation}

Equation~\eqref{eq:DT-compact} is an algebraic rewriting of Georgiev's Eq.~(46). For $0<C<1$, it diverges at $\vartheta=k\pi$, where the signal is extremal. For $C=1$, the numerator and denominator cancel away from those points, yielding $1/\omega$, as in his Eq.~(48); exactly at an extremum, the defining ratio is $0/0$ and only its interior limit is finite. These algebraic facts are not in dispute. The issue is their interpretation.

\section{The time scale is an inverse Fisher sensitivity}
\label{sec:fisher}
The starting point of our analysis is Georgiev's Eq.~(51) \cite{Georgiev2024}, which expresses the instantaneous standard deviation of an observable $\hat A$ in terms of the rate of change of its expectation value and the characteristic time $\Delta T(t)$,
\begin{equation}
\Delta\hat A(t)=\left|\frac{\dd}{\dd t}\langle\hat A\rangle\right|(t)\,\Delta T(t).
\label{eq:Georgiev51}
\end{equation}
Immediately after Eq.~\eqref{eq:Georgiev51}, Georgiev interprets $\Delta T(t)$ as an instantaneous ``hypothetical time duration that would have been taken by the quantum observable moving with a constant speed equal to the instantaneous rate of change $\left|\frac{\dd}{\dd t}\langle\hat A\rangle\right|(t)$ until a distance equal to the instantaneous standard deviation $\Delta \hat A(t)$ would have been traversed.'' \cite[p.~9]{Georgiev2024} Because the signal accelerates and decelerates, he concludes that $\Delta T(t)$ does not disclose how well the system measures the passage of time, in contrast with the constant $\Delta E$. We now apply the estimation route stated above to show why that conclusion is too strong.

A projective measurement of $\sx$ returns $m=\pm1$, not the expectation value $\langle\sx\rangle$ itself. Georgiev's projector probabilities in Eqs.~(39)--(40) can be written as
\begin{equation*}
  p_m(t)=\frac{1}{2}
  +m\!\left[
  \operatorname{Re}(\alpha_1\alpha_2^*)\cos(\omega t)
  +\operatorname{Im}(\alpha_1\alpha_2^*)\sin(\omega t)
  \right]
  =\frac{1+m \langle\sx\rangle}{2},
  \qquad m=\pm1.
  \label{eq:probabilities}
\end{equation*}

Whenever $0<p_m<1$, the classical Fisher information per projective measurement about the parameter $t$ is
\begin{align}
 F_x(t)
 &=\sum_{m=\pm1}
 \frac{[\ddt p_m(t)]^2}{p_m(t)}
 =\frac{[\ddt\ssx]^2}{1-\ssx^2}
 \notag\\
 &=\frac{\omega^2C^2\sin^2\vartheta}
 {1-C^2\cos^2\vartheta}.
 \label{eq:Fx}
\end{align}

The Fisher information quantifies the local statistical distinguishability of the outcome distributions at neighbouring values of $t$ \cite{Paris2009}. A large $F_x(t)$ means that a small change of time appreciably changes the probabilities $p_m(t)$; $F_x(t)=0$ means that this selected readout is insensitive to time to first order. The Fisher information is calculated from the time-dependent outcome probabilities of a specified measurement. It is therefore a property of the statistical readout model $p_m(t)$, not the expectation value of a quantum observable and not, by itself, the variance of a time estimator.

At a locally identifiable operating point, where $F_x(t)>0$, a binary Pauli outcome obeys $\Delta\sx=\sqrt{1-\ssx^2}$ and Eqs.~\eqref{eq:DT-compact} and \eqref{eq:Fx} give the exact identity
\begin{equation}
  \boxed{\Delta T(t)=\frac{1}{\sqrt{F_x(t)}}}.
  \label{eq:identity}
\end{equation}

Thus Georgiev's Eq.~(51) has a direct statistical meaning for his chosen readout: $\Delta T$ is its inverse local sensitivity to a change of $t$. It is neither the variance of a time operator nor the elapsed duration of a single run. Its phase dependence instead describes how the information about $t$ available from the $\sx$ measurement changes during the spin precession.

At a submaximal-coherence extremum, $0<C<1$ and $\vartheta=k\pi$, one has $\ddt p_m=F_x=0$. The divergence $\Delta T=\infty$ expresses loss of first-order sensitivity; it is not an infinite variance of physical time. The probabilities of the $\sx$ outcomes change only to second
order in a small displacement from the extremum and are identical
for opposite displacements.

Georgiev is nevertheless right that $\Delta E$ has a different status. To make this distinction precise, we introduce the quantum Fisher information $F_Q$. At a fixed value of $t$, $F_Q$ is the maximum classical Fisher information obtainable from the time-dependent quantum state by optimizing over all possible measurements. It therefore quantifies the greatest local sensitivity to $t$ available in the state before a particular readout is selected \cite{BraunsteinCaves1994,BraunsteinCavesMilburn1996}.

For the pure state undergoing Georgiev's unitary evolution,
\begin{equation}
  F_Q
  =4\left(
  \langle\dot\Psi|\dot\Psi\rangle
  -|\langle\Psi|\dot\Psi\rangle|^2
  \right)
  =\frac{4(\Delta\Ham)^2}{\hbar^2}
  =\omega^2C^2
  \label{eq:FQ}
\end{equation}
where
$|\dot\Psi\rangle\equiv
\frac{\dd}{\dd t}|\Psi(t)\rangle$.

For mixed states undergoing unitary evolution, one generally has $F_Q\leq4(\Delta\Ham)^2/\hbar^2$, and the inequality can be strict.
Returning to Georgiev's pure-state model, at regular operating points one finds

\begin{equation}
 F_Q-F_x
 =
 \frac{\omega^2C^2(1-C^2)\cos^2\vartheta}
 {1-C^2\cos^2\vartheta}
 \geq0.
 \label{eq:difference}
\end{equation}
The nonnegative difference $F_Q-F_x$ quantifies the local timing information available in the quantum state but not accessed by the selected $\sx$ measurement. When $F_x=F_Q$, the measurement is locally optimal; when $F_x=0<F_Q$, the state remains sensitive to $t$, but the $\sx$ measurement fails to reveal that sensitivity to first order.

Consequently, at regular operating points with $F_x(t)>0$,
\begin{equation}
 \Delta T\,\Delta\Ham
 =\frac{\hbar}{2} \frac{\sqrt{1-C^2 \cos^2\vartheta}}{|\sin\vartheta|}
 =\frac{\hbar}{2}\sqrt{\frac{F_Q}{F_x}}
 \geq\frac{\hbar}{2}.
 \label{eq:mt-fisher}
\end{equation}

Thus the Mandelstam--Tamm inequality is the information hierarchy $F_x\leq F_Q$: the selected readout cannot contain more information about the time translation than the state itself. In this precise sense, the constant $\Delta E$ fixes the best sensitivity allowed by the state, while the phase-dependent $\Delta T$ describes how much of that sensitivity Georgiev's selected $\sx$ measurement actually accesses. The two quantities answer different questions, so the variability of one does not make it meaningless.

\begin{figure}[t]
  \centering
  \includegraphics[width=1\linewidth]{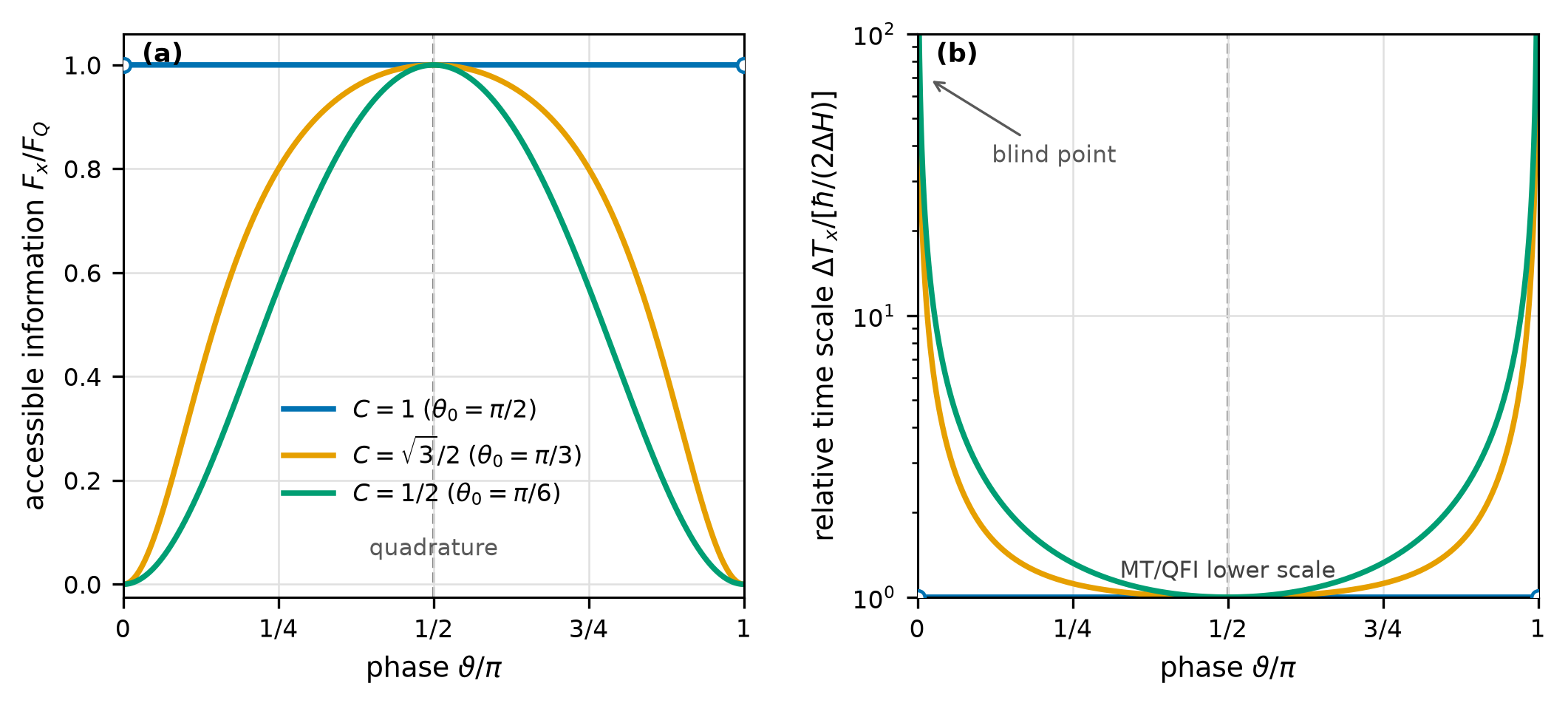}
  \caption{\textbf{Temporal sensitivity of Georgiev's $\sx$ clock.} We take $\alpha_1\alpha_2^*$ real and positive, so that $\vartheta=\omega t$. Panel (a) shows the accessible fraction $F_x/F_Q$; panel (b) shows $\Delta T/[\hbar/(2\Delta\Ham)]=\sqrt{F_Q/F_x}$. At $\omega t=\pi/2$, the maximum-slope (quadrature) point, the $\sx$ readout attains the quantum Fisher-information bound $F_x=F_Q$ for every $C>0$. For $C<1$, $F_x$ vanishes at the extrema and the relative time scale diverges. Open circles for $C=1$ denote interior limits at the nonregular endpoints.}
  \label{fig:fisher}
\end{figure}

For $N$ independently and identically prepared states, Fisher
information adds. At a locally identifiable regular point with
$F_x(t)>0$, any locally unbiased estimator therefore obeys
\begin{equation}
  \Var(\hat t)
  \geq
  \frac{1}{N F_x(t)}
  =
  \frac{\Delta T^2}{N}.
  \label{eq:crb}
\end{equation}
Thus $\Delta T/\sqrt{N}$ is the local Cram\'er--Rao sensitivity scale, which can be approached asymptotically by an efficient estimator under the usual regularity conditions. The corresponding statement about estimation uncertainty is therefore expressed by Eq.~\eqref{eq:crb}, not by $\Delta T$ alone. Finite samples, estimator bias, and global ambiguities arising from the periodicity of the signal require separate treatment \cite{Tsang2012,Meyer2025}. Figure~\ref{fig:fisher} summarizes the accessible information and the corresponding relative time scale over one half-period.

\section{Divergences and Georgiev's tick--tock argument}
\label{sec:divergence}

Georgiev draws his strongest conclusion after Eqs.~(52)--(54). At a maximum or minimum of $\langle\sx\rangle$, $\Delta T$ diverges for $C<1$, whereas the interval between consecutive extrema is exactly $\delta t=\pi/\omega$. He argues that $\Delta T$ would therefore ``wrongly predict'' an infinite standard deviation of time and concludes that it is not a meaningful representation of time. This comparison puts together two different quantities.

For $0<C<1$ at $\vartheta=k\pi$, $F_x=0$ while $F_Q>0$. The divergence of $\Delta T$ means that the selected $\sx$ measurement has no first-order sensitivity to a small time displacement from the extremum. It does not imply that external time has a statistical variance, nor does it make the known period $\pi/\omega$ uncertain. Estimation exactly there is nonregular because the leading change of the $\sx$ outcome probabilities is quadratic and does not determine on which side of the extremum the system lies.

The complementary readout has
\begin{equation*}
 F_y(t)=
 \frac{\omega^2C^2\cos^2\vartheta}
 {1-C^2\sin^2\vartheta}.
 \label{eq:Fy}
\end{equation*}
At an extremum of $\ssx$, $F_y=F_Q$; conversely, the $\sx$ readout is optimal at $\cos\vartheta=0$ for every $C>0$. Thus the divergence identified by Georgiev is readout-specific, not a disappearance of temporal information from the state. Operationally, a clock can switch to the complementary quadrature near an $\sx$ extremum; without prior phase knowledge, the same idea requires a local or adaptive measurement strategy.

Georgiev's Eqs.~(52)--(54) relate the energy gap to the separation between consecutive ``tick'' and ``tock'' extrema:
\begin{align*}
 \omega&=\frac{|E_1-E_2|}{\hbar},
 \notag\\
 \delta t&=|t_2-t_1|=\frac{\pi}{\omega},
 \notag\\
 \delta E\,\delta t&=\frac{h}{2},
 \qquad
 \delta E=|E_1-E_2|.
 \label{eq:Georgiev-ticktock}
\end{align*}

These relations fix the separation between ideal extrema of a signal whose frequency is already known. They do not determine the statistical precision with which an extremum can be located from measurement outcomes. For $C<1$, an extremal expectation value is $\ssx=\pm C$, whereas one projective outcome remains random, $m=\pm1$. Locating a tick therefore requires an ensemble of identically prepared spins, repeated runs, or prior synchronization. Even for $C=1$, the endpoint model is nonregular and the $\sx$ readout does not distinguish the two signs of a displacement. Periodicity also leaves the cycle number undetermined without coarse timing information. Hence Georgiev's Eq.~(53) is fully compatible with a divergent local $\sx$-readout scale: the former is a calibrated half-period, while the latter describes the loss of first-order timing information in a specified measurement.

\section{Conclusions}

Georgiev is correct that $\Delta T$ is neither a time-operator variance nor an elapsed time produced by a single run. His stronger inference does not follow. In his own $\sx$ experiment, $\Delta T$ is the inverse Fisher sensitivity at regular operating points. Its phase dependence describes the performance of the selected readout, and its divergence identifies a locally blind quadrature; neither conflicts with the exactly calibrated half-period $\delta t=\pi/\omega$. A complementary quadrature can retain the information that the $\sx$ readout loses. These qualifications preserve Georgiev's conceptual warning while replacing physical meaninglessness with a precise, measurement-dependent operational meaning.

\funding{This work was supported by the National Science Centre, Poland (grant no. 2024\slash 53\slash N\slash HS1\slash 03316).}

\dataavailability{No external dataset was used in this study.}

\printbibliography[title={References}]

\end{document}